\documentclass[11pt,superscriptaddress,aps,prd,preprint]{revtex4}
\usepackage[active]{srcltx}
\usepackage{graphicx}
\usepackage[utf8]{inputenc}
\usepackage{amsmath}
\usepackage{verbatim}
\usepackage{xcolor}
\usepackage{subcaption}
\usepackage{times,txfonts}
\usepackage{float}%para fixar as figuras na posição desejada

\def\hyper#1#2#3#4{{}_2F_1 \left(#1\text{ , }#2\text{ ; }#3\text{ ; }~#4\right)}

\begin{document}

\title{Horizon Fractalization in Black Strings Ungravity}

\author{I. D. D. Carvalho}
\affiliation{Universidade Federal do Ceará, Fortaleza, Ceará, Brazil}
\email{icarodias@alu.ufc.br}

\author{J. Furtado}
\affiliation{Universidade Federal do Cariri, Juazeiro do Norte, Ceará, Brazil}
\email{job.furtado@ufca.edu.br}

\author{R. R. Landim}
\affiliation{Universidade Federal do Ceará, Fortaleza, Ceará, Brazil}
\email{renan@fisica.ufc.br}

\author{G. Alencar}
\affiliation{Universidade Federal do Ceará, Fortaleza, Ceará, Brazil}
\email{geova@fisica.ufc.br}

\date{\today}

\begin{abstract}

In this paper we study the scalar(tensor) and vector unparticle corrections for cosmic and black strings. Initially we have considered an static cosmic string ansatz from which we obtain the solution in terms of first and second kind Bessel functions. We have also obtained the solution for black string in the unparticle scenario. We could identify two regimes, namely, a gravity dominated regime and an ungravity dominated regime.  In the gravity dominated regime the black string solution recovers the usual solution for black strings. The Hawking temperature was also studied in both regimes and in the ungravity dominated regime. As in the static and rotating black hole,  we found a fractalization of the event horizon. This points to the fact that fractalization is a natural consequence of unparticles. Finally, we study the thermodynamic of the black string in the ungravity scenario by computing the entropy, heat capacity and free energy. For both cases we find that, depending on the region of the parameter $d_U$, we can have phase transitions.
  
\end{abstract}

\maketitle

\section{Introduction}\label{Sec-1}

The unusual properties of non-trivial scale invariance of matter in the IR regime was first studied by Banks and Zaks \cite{Banks:1981nn}. Georgi suggested \cite{Georgi:2007ek, Georgi:2007si} a coupling between the standard model (SM) fields and this scale invariant sector of a high energy physics described by the Banks-Zaks (BZ) fields. The interactions between the standard model fields and the BZ fields are mediated by the exchange of highly massive particles $M_U$, obeying the scale-supressed non-renormalizable Lagrangian density
\begin{equation}
\mathcal{L}=\frac{1}{M_U^k}\mathcal{O}_{SM}\mathcal{O}_{BZ},
\end{equation}
where $\mathcal{O}$ stands for the field operators of dimension $d_{SM}$ and $d_{BZ}$, while $k=d_{SM}+d_{BZ}-4$ in order to guarantee a dimensionless action. It was argued by Georgi \cite{Georgi:2007si} that below some scale energy limit ($\Lambda_U$) the interactions between the standard model and the hidden sector become stronger. As the scale-invariance is manifested, massive particles are not allowed and this sector was dubbed ``unparticles'', since the notion of mass in this sector is meaningless.

Since the proposal of the unparticle physics several papers were published in particle physics \cite{Aliev:2017bme, Bagheri:2017roa, VanSoa:2018pkb}, condensed matter analogue models \cite{Limtragool:2016ghz, LeBlanc:2014xpa}, extra dimensions \cite{Soa:2019whl} among others. A very interesting work by Gaete and Spallucci \cite{Gaete:2008wg} study the unparticle dynamics in a standard quantum field theory scenario, and they construct effective actions for scalar, gravitation and vector gauge unparticles. Also in cosmological and astrophysical scenarios, several were carried out including studies with black holes \cite{Gaete:2010sp, Mureika:2010je, Lee:2011wc, Alencar:2018vvb} and white dwarfs \cite{Moussa:2021fsb, deSouza:2012ap}. One of these black holes works draws the attention to the fact that the exterior event horizon presents a surface with fractal dimension equals to $d_U$ \cite{Gaete:2010sp}. Recently, some of the present authors generalized the work of \cite{Gaete:2010sp} to rotating black holes and found that the the fractalization is also present\cite{Alencar:2018vvb}. It is also noticeable that a recent paper on Casimir effect with unparticles shows a fractalization in the parallel plates dimension \cite{Frassino:2013lya}. The dimensional reduction seems to be a natural behavior in high energy  and its comprehension in fundamental at this regime scales\cite{Stojkovic:2013xcj,Carlip:2017eud}.
 However, as far as we know, no studies were carried out in order to investigate the influence of the unparticle physics in cosmic strings and black strings.

Objects such as cosmic strings may have been formed due to phase transitions in the early universe \cite{Vilenkin}. The spacetime geometry related to an infinitely long and straight cosmic string is characterized by a planar angle deficit on the two-surface orthogonal to the string. In addition to that, the spacetime is flat, locally, except on the top of the string. Cosmic strings can also be described in the realm of a classical field theory when the energy-momentum tensor related to the vortex configuration of the Higgs-Maxwell system \cite{Nielsen:1973cs} couples to the Einstein equations. 

On the other hand, black strings are higher dimensional solutions of the Einstein equations in D-dimensional spacetime \cite{Duff:1987cs}. These vacuum objects can be understood as symmetric generalizations of black holes under translations being the event horizon topologically equivalent to $S_2 \times R$ (or $S_2 \times S_1$ in the case of black rings) and spacetime is asymptotically $M_{D-1} \times S_1$ for a zero cosmological constant or $AdS_{D-1} \times S_1$ for a negative one \cite{Emparan:2001wn, Bellucci:2010gb}. A black string is a particular case of a black p-brane solution (with $p = 1$), where $p$ is the number of extra spatial dimensions to ordinary space. Hence a p-brane gives rise to a (p + 1)-dimensional world-volume in spacetime. In string theory, a black string is described by a D1-brane surrounded by a horizon. It is widely known that since an open string propagates through spacetime, it is necessary that its endpoints to lie on a D2-brane on which it satisfies Dirichlet boundary condition, and the dynamics on the D-brane worldvolume is described by a gauge field theory \cite{Polchinski:1995mt, Polchinski:1994fq}. In some models our universe is understood as a higher dimensional world-brane, thus the gravitational collapse of galactic matter would produce black holes lying on the brane. Black string solutions are nevertheless unstable to long wavelength perturbations (at least in asymptotically flat spaces) since the localized black hole is entropically preferred to a long segment of string. The string’s horizon therefore has a tendency to form a line of black holes. However, in AdS the space acts like a confining box which prevents fluctuations of long wavelengths from developing. For instance, in a Randall Sundrum domain wall in a five dimensional AdS spacetime the black string is stable at least far from the AdS horizon. Near the horizon, the string is likely to pinch off to become a stable shorter cigar-like singularity \cite{Chamblin:1999by}.

In this paper we study the unparticle corrections for cosmic and black strings. This paper is organized as follows: In the next section we present a brief review of the unparticle theory. In section III we study the unparticle corrections for cosmic strings and in section IV we obtain the black string solution in the unparticle scenario. In section V we study the thermodynamic properties of black strings in the unparticle physics domain and in section VI we draw our conclusions. 

%%%%%%%%%%%%%%%%%%%%%%%%%%%%%%%%%%%%%%%%%%%%%%%%%%%%%%%%%%%%%%%%%%%%%%%%%%%%%%%%%%%

\section{Unparticle Theory}\label{Sec-2}
The un-gravity action is \cite{Gaete:2010sp}
\begin{eqnarray}\label{1}
    S_u = \frac{1}{2\kappa^2} \int d^4x \sqrt{-g} \left[1 + \frac{A_{d_U}}{(2d_U -1)\sin{(\pi d_U)}} 
    \frac{\kappa_{*}^{2}}{\kappa^{2}}\left(-\frac{D^2}{\Lambda^2_U}\right)^{1-d_U}\right]^{-1} \mathcal{R},
\end{eqnarray}
where $\kappa^2 = 8\pi G$, $\kappa_{*}^2$ is represents the un-gravitational Newton constant, and $ A_{d_U} = 16\pi^{5/2} \Gamma(d_U + 1/2)\left[(2\pi)^{2d_U} \Gamma(2d_U) \Gamma(d_U-1)\right]^{-1} $  \cite{Gaete:2010sp}, and $D^2$ is the generally covariant D'alambertian. The Einstein Field Equations (EFE) is given by
\begin{eqnarray}\label{2}
    R^\mu_\nu - \frac{1}{2} \delta^\mu_\nu \mathcal{R} &=& \kappa^2 T^\mu_\nu+ \frac{\kappa_*^2A_{d_U}}{\sin{(\pi d_U)}}\frac{\Lambda_U^{2 - 2d_U}}{(2d_U -1)}\left(-\frac{1}{D^2}\right)^{1-d_U} T^\mu_\nu \nonumber \\
    &=& \kappa^2 T^\mu_\nu + \frac{\kappa_*^2A_{d_U}}{\sin{(\pi d_U)}} {T_U}^\mu_\nu
\end{eqnarray}
The second term of right side of EFE is the unparticle correction of energy-momentum tensor. In Ref. \cite{Mureika:2010je} it was found an exact solution of EFE for vector particles, and this solution is Reissner-Nordstrom like. In this paper, we'll consider an axially symmetric \textit{ansatz} for the metric, $(g_{\mu\nu})=\text{diag}\left(g_{tt}(r),g_{rr}(r),g_{\phi\phi}(r),g_{zz}(r)\right)$, since we are interested in studying solutions of Cosmic String and Black String with unparticle corrections. Moreover, we adopt the metric's signature $(-1,1,1,1)$, so  we consider the energy-momentum tensor as $(T^\mu_\nu) = \text{diag}(-\rho,p,p_\phi,p_z)$, which can be represented by
\begin{eqnarray}\label{energy-momentum}
    T^\mu_\nu = (\rho + p_\phi) u^\mu u_\nu + (p-p_\phi)w^\mu w_\nu+ (p_z-p_\phi)Z^\mu Z_\nu + p_\phi~\delta^\mu_\nu,
\end{eqnarray}
where $(u_\mu) = \left(\sqrt{-g_{tt}},0,0,0\right)$ is the four-velocity field, $(w_\mu) = \left(0,\sqrt{g_{rr}},0,0\right)$ and $(Z_\mu) = \left(0,0,0,\sqrt{g_{zz}}\right)$ are  units spacelike vectors.

Let us consider a static string which has a density of energy $\rho(r) =\mu(2\pi r)^{-1} \delta(r)= \mu \delta(x)\delta(y)$, where $\mu$ is the linear density of mass. We can compute the contribution of unparticle in equation (\ref{2}) by calculating the term $T_U$. We achieve that by switching to isotropic (free-falling), Cartesian-like coordinate for computational convenience,
\begin{eqnarray}\label{6}
    \left(-\frac{1}{D^2}\right)^{1-d_U} T^0_0 &=& - \mu \left(-\frac{1}{\nabla_{\vec{X}}^2}\right)^{1-d_U} \delta \left(\vec{X}\right) \nonumber\\
    &=& - \mu \left(-\frac{1}{\nabla_{\vec{X}}^2}\right)^{1-d_U} \frac{1}{(2\pi)^2}\int d^2k e^{i \vec{k}\cdot\vec{X}} \nonumber \\
    &=&- \mu \frac{1}{(2\pi)^2}\int d^2k \left(\frac{1}{k^2}\right)^{1-d_U} e^{i \vec{k}\cdot\vec{X}},
\end{eqnarray}
where $\vec{X}$ is a vector in a direction perpendicular to the string. Hence, by applying the Schwinger representation for $(1/k^2)^{1-d_U}$, we obtain 
\begin{eqnarray}\label{7}
    \left(\frac{1}{k^2}\right)^{1-d_U} = \frac{1}{\Gamma(1-d_U)}\int_0^\infty ds s^{-d_U} e^{-sk^2}.
\end{eqnarray}
Yielding for the equation (\ref{6})
\begin{eqnarray}\label{8}
    \left(-\frac{1}{D^2}\right)^{1-d_U} T^0_0 = - \mu \frac{1}{(2\pi)^2}\int d^2k \left[\frac{1}{\Gamma(1-d_U)}\int_0^\infty ds s^{-d_U} e^{-sk^2}\right] e^{i \vec{k}\cdot\vec{X}}.
\end{eqnarray}
Finally, when we compute this integration, we find
\begin{eqnarray}\label{9}
    \left(-\frac{1}{D^2}\right)^{1-d_U} T^0_0 = - \mu \frac{2^{2d_U - 2}}{\pi} \frac{\Gamma(d_U)}{\Gamma(1-d_U)} \left(\frac{1}{\vec{X}^2}\right)^{d_U},
\end{eqnarray}
which allow us to write a proper expression for the energy density of a static string with unparticle as
\begin{eqnarray}\label{density0}
    \rho(r) = \frac{\mu}{2\pi}\frac{\delta(r)}{r} + \mu\left(\frac{\kappa^{2}_{*}}{\kappa^2}\right) \frac{ A_{d_U}}{\sin{\pi d_U}}\frac{\Lambda^{2-2d_U}_U}{(2d_U-1)} \frac{2^{2d_U-2}}{\pi}\frac{\Gamma(d_U)}{\Gamma(1-d_U)} \left(\frac{1}{r}\right)^{2d_U}.
\end{eqnarray}
When we replace $ A_{d_U} = 16\pi^{5/2} \Gamma(d_U + 1/2)\left[(2\pi)^{2d_U} \Gamma(2d_U) \Gamma(d_U-1)\right]^{-1}$ in equation (\ref{density0}), we find
\begin{eqnarray}\label{density}
    \rho(r) = \frac{\mu}{2\pi}\left[\frac{\delta(r)}{r}+\frac{R^{2d_U-2}}{r^{2d_U}}\right],
\end{eqnarray}
where
\begin{eqnarray}\label{unradiu}
    R = \left[\left(\frac{\kappa_*^2}{\kappa^2}\right)\frac{8\Lambda_U^{2 - 2d_U}}{\sin{(\pi d_U)}(2d_U -1)\pi^{2d_U-5/2}}\frac{ \Gamma(d_U)\Gamma(d_U + 1/2)}{\Gamma(1-d_U) \Gamma(2d_U) \Gamma(d_U-1)}\right]^\frac{1}{2d_U-2}.
\end{eqnarray}

Also it is important to highlight that the $R$ parameter sets a proper unparticle length scale so that, similarly to the authors in \cite{Gaete:2010sp, Alencar:2018vvb}, it is possible to distinguish between two regimes. The gravity dominated (GD) regime is defined by $R<<r$ and the ungravity dominated (UGD) regime by $R>>r$. Note that in the GD regime the second term of (\ref{density}) vanishes and we recover the usual cosmic string energy density.

\begin{figure}[h]
    \centering  %para centralizarmos a figura
    \includegraphics[width=10cm]{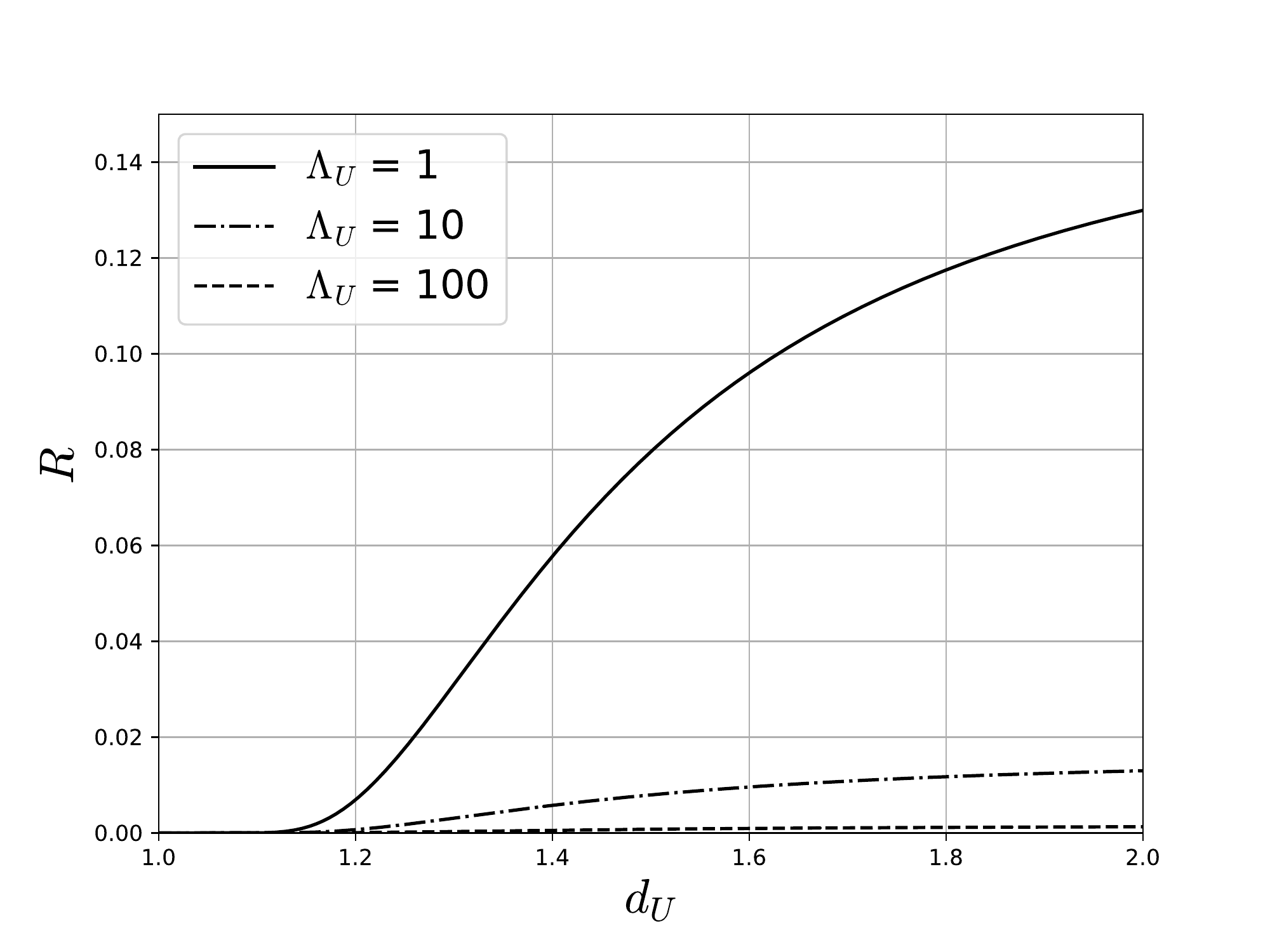}
    \caption{Behavior of $R$ for different $\Lambda_U$ as function of $d_U$, with $(\kappa_{*}/\kappa)^2=0.5$}
    \label{fig-beta-constant}
\end{figure}

In the next section, we will present the solution of static cosmic string with  unparticle corrections.

%%%%%%%%%%%%%%%%%%%%%%%%%%%%%%%%%%%%%%%%%%%%%%%%%%%%%%%%%%%%%%%%%%%%%%%%%%%%%%%%%%%
\section{Cosmic String with unparticle corrections}
 In this section, we will study the spacetime solution which is static and has cylindrical symmetry. Consider the \textit{ansatz} of the line-element as follows 
 \begin{eqnarray}\label{3}
     ds^2 = -dt^2 + dr^2 + b(r)^2 d\phi^2 + dz^2.
 \end{eqnarray}
 The components of the Ricci tensor are $R_{\mu\nu}=R^\alpha_{\mu\alpha\nu} = \partial_{[\alpha}\Gamma^{\alpha}_{\nu]\mu}+ \Gamma^{\alpha}_{\beta[\alpha}\Gamma^{\beta}_{\nu]\mu}$, where $V_{[\mu\nu]}= V_{\mu\nu}-V_{\nu\mu}$. The Einstein Tensor is
 \begin{eqnarray}\label{4}
     G^\mu_\nu = \frac{1}{b(r)}\frac{d^2b(r)}{dr^2} ~\text{diag}\left(1,0,0,1\right).
 \end{eqnarray}
We know that the $T^0_0 = - \rho(r)$, so we can see by EFE $G^\mu_\nu = \kappa^2 T^\mu_\nu$ that $T^\mu_\nu = -\rho(r) \text{diag}(1,0,0,1)$. Then the EFE resumes to 
\begin{eqnarray}\label{5}
    \frac{1}{b(r)}\frac{d^2b(r)}{dr^2} = - \kappa^2 \rho(r).
\end{eqnarray}
This equation can be rewritten as
\begin{eqnarray}\label{12}
    \frac{d^2b(r)}{dr^2} +\frac{\mu\kappa^2}{2\pi}\left[ \frac{\delta(r)}{r} + R^{2d_U-2} r^{-2d_U}\right] b(r) = 0.
\end{eqnarray}

\begin{figure}[h]
    \centering  %para centralizarmos a figura
    \includegraphics[width=10cm]{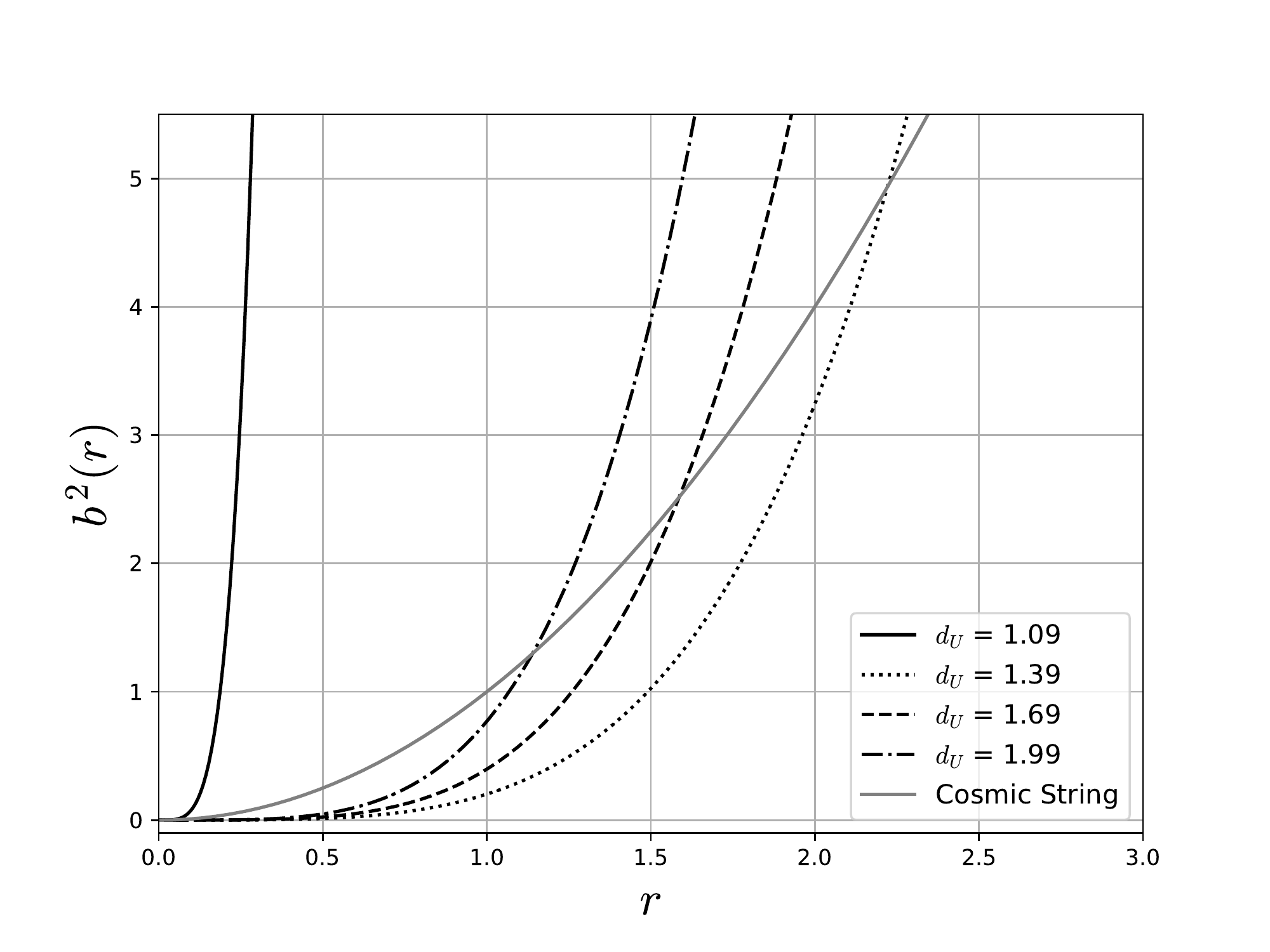}
    \caption{Metric component $g_{\phi\phi}$ of the unparticle static cosmic string as a function of the radial coordinate $r$ for some values of $d_U$ , with $\Lambda_U = 10$, $\mu = 0.7$, $\kappa_* = 10^3$}
    \label{metric-component-phi}
\end{figure}

Now, we need to properly solve the equation (\ref{12}). Let us perform the following variable transformation $b(r) = \sqrt{r}f(r)$, such that the equation (\ref{12}) becomes
\begin{eqnarray}\label{13}
    \left\{\frac{d^2f}{dr^2}+ \frac{1}{r}\frac{df}{dr}+\left[\frac{\mu\kappa^2}{2\pi} \frac{\delta(r)}{r} + \frac{\mu\kappa^2}{2\pi}R^{2d_U-2}~ r^{-2d_U}- \frac{1}{4r^2}\right]f\right\} \sqrt{r} = 0.
\end{eqnarray}
First, we need to solve this equation for $r>0$. After we impose that when $R/r \rightarrow 0$ we must find cosmic string spacetime. When we consider $r>0$ the equation (\ref{13}) simplifies to
\begin{eqnarray}\label{14}
    \frac{d^2f}{dr^2}+ \frac{1}{r}\frac{df}{dr}+\left[\frac{\mu\kappa^2}{2\pi}R^{2d_U-2} r^{-2d_U}- \frac{1}{4r^2}\right]f = 0.
\end{eqnarray}
Now, we can perform a transformation of variable $ x = \sqrt{\frac{\mu\kappa^2}{2\pi}}\frac{R^{d_U-1}}{\left(d_U-1\right)} r^{1-d_U}$, allowing the equation (\ref{14}) to be written as 
\begin{eqnarray}\label{15}
    \frac{d^2f}{dx^2}+ \frac{1}{x}\frac{df}{dx}+\left[1- \frac{\left(\frac{1}{2du-2}\right)^2}{x^2}\right]f = 0.
\end{eqnarray}
This is the Bessel differential equation. The solution of equation (\ref{15}) depends on $\left(\frac{1}{2du-2}\right)$. When we define $\xi \equiv 1/(2du-2)$ as a positive real number the equation (\ref{15}) becomes
\begin{eqnarray}\label{16}
    \frac{d^2f}{dx^2}+ \frac{1}{x}\frac{df}{dx}+\left[1- \frac{\xi^2}{x^2}\right]f = 0,
\end{eqnarray}
whose solution is 
\begin{eqnarray}\label{17}
    f(x) = C_1 J_\xi (x) + C_2 N_\xi (x).
\end{eqnarray}
Here $J_\xi(x)$ and $N_\xi(x)$ are the first and second kind Bessel function of order $\xi$, respectively.
Let us present some considerations regarding the boundary conditions. Initially, we want that when $R/r \rightarrow 0$, $b(r) = (1-8G\mu)^{1/2}r$, that is, the spacetime becomes the cosmic string spacetime. Since $x$ is proportional to $(R/r)^{d_U-1}$ and $b(r)=\sqrt{r}f(x(r))$, we can study the equation (\ref{17}) when $x \ll 1$. In regime $x \ll 1$ and $\xi > 0$, equation (\ref{17}) can be written as
\begin{eqnarray}\label{18}
    f(x)= C_1\frac{1}{\Gamma{(\xi+1)}}\left(\frac{x}{2}\right)^\xi - C_2 \frac{\Gamma(\xi)}{\pi}\left(\frac{x}{2}\right)^{-\xi}.
\end{eqnarray}
%%%%%%%%%%%%%%%%%%%%%%%%%%%%%%%%%%%%%%%%%%%%%%%%%%%%%%%%%%%%%%%%%%%%%%%%%%%%%%%%%%%%%%%%%%
Replacing $ x = \sqrt{\frac{\mu\kappa^2}{2\pi}}\frac{R^{d_U-1}}{\left(d_U-1\right)} r^{1-d_U}$ in equation (\ref{18}), when we impose the limit $R/r \ll 1$ we shall find $f(r) = (1-8G\mu)^{1/2}\sqrt{r}$. Hence $C_1 = 0$ and
\begin{eqnarray*}
C_2 = -(1-8G\mu)^{1/2}\frac{\pi}{\Gamma{(\xi)}}\left(\xi \sqrt{\frac{\kappa^2\mu}{2\pi}}\right)^\xi R^{1/2}.
\end{eqnarray*}
Consequently, the component $g_{\phi\phi}$ of the metric is
\begin{eqnarray}\label{21}
    b^2(r) = (1-8G\mu)r^2\left[\frac{\pi^2}{\Gamma^2{(\xi)}}\xi^{2\xi}\left( 4G\mu\right)^{\xi} \frac{R}{r}\right]  {N_\xi}^2 \left(4\xi\sqrt{G\mu}\left(\frac{R}{r}\right)^{\frac{1}{2\xi}}\right).
\end{eqnarray}
The metric component $g_{\phi\phi} = b^2(r)$, the behavior of this component is given at Figure \ref{metric-component-phi}.

In order to understand the structure of this spacetime, that is, how the unparticles modify the cosmic string spacetime, We must analyze the curvature scalars. Ricci's scalar for the ansatz $ds^2 = -dt^2 + dr^2 + b^2(r)d\phi^2 + dz^2$ is $\mathcal{R} = - \frac{2}{b}\frac{d^2b}{dr^2}$. Then Ricci's scalar for our solution is given by 
\begin{eqnarray}\label{22}
    \mathcal{R}=\frac{\mu\kappa^2}{\pi}\left[ \frac{\delta(r)}{r} + \frac{1}{r^2}\left(\frac{R}{r}\right)^{2d_U-2}\right],
\end{eqnarray}
where we used equation (\ref{13}).
%%%%%%%%%%%%%%%%%%%%%%%%%%%%%%%%%%%%%%%%%%%%%%%%%%%%%%%%%%%%%%%%%%%%%%%%%%%%%%%%%%%

\section{Black string with unparticle corrections}

Let us consider the following line element for the black string 
\begin{eqnarray}\label{23}
    ds^2=-f(r)dt^2+\frac{1}{f(r)}dr^2+r^2d\phi^2+\alpha^2r^2dz^2,
\end{eqnarray}
where $t\in(-\infty, \infty)$, the radial coordinate $r\in[0,\infty)$, the angular coordinate $\phi\in[0,2\pi)$ and the axial coordinate $z\in(-\infty,\infty)$. The $\alpha$ parameter is considered as $\alpha^2=-\Lambda/3$.

For the black string in the un-gravity scenario, the Einstein-Hilbert effective action is composed by (\ref{1}) added by the cosmological constant contribution, i.e., 
\begin{eqnarray}\label{24}
    S_u = \frac{1}{2\kappa^2} \int d^4x \sqrt{-g} \left\{\left[1 + \frac{A_{d_U}}{(2d_U -1)\sin{(\pi d_U)}} 
    \frac{\kappa_{*}^{2}}{\kappa^{2}}\left(-\frac{D^2}{\Lambda^2_U}\right)^{1-d_U}\right]^{-1} \mathcal{R}-2\Lambda\right\}.
\end{eqnarray}
The Einstein equation in (\ref{2}) remains valid, however the left-hand-side will be modified due to the different ansatz for the metric in (\ref{23}). The EFE for this ansatz are
\begin{eqnarray}
    G^t_t -3\alpha^2 &=& \frac{1}{r} \frac{df(r)}{d r}  + \frac{f(r)}{r^2} - 3\alpha^2, \label{25}\\
    G^r_r -3\alpha^2 &=& \frac{1}{r} \frac{df(r)}{d r}  + \frac{f(r)}{r^2} - 3\alpha^2, \label{26}\\
    G^\phi_\phi -3\alpha^2 &=& \frac{1}{2}\frac{d^{2}f(r)}{d r^{2}}+ \frac{1}{r} \frac{df(r)}{d r} - 3\alpha^2, \label{27}\\
    G^z_z -3\alpha^2 &=&\frac{1}{2}\frac{d^{2}f(r)}{d r^{2}}+ \frac{1}{r} \frac{df(r)}{d r} - 3\alpha^2.\label{28}
\end{eqnarray} 
We can see that the energy-momentum tensor for the ansatz of equation (\ref{23}) is $T^\mu_\nu = -\rho(r)\text{ diag}(1,1,0,0) + p_l(r)\text{ diag}(0,0,1,1)$, where $p_l = p_\phi = p_z$. This way we can find $f(r)$ by solving $G^t_t -3\alpha^2 = -\kappa^2 \rho(r)$, we find

\begin{figure*}[h!]
    \centering
\begin{subfigure}{.5\textwidth}
  \centering
  \includegraphics[scale=0.8]{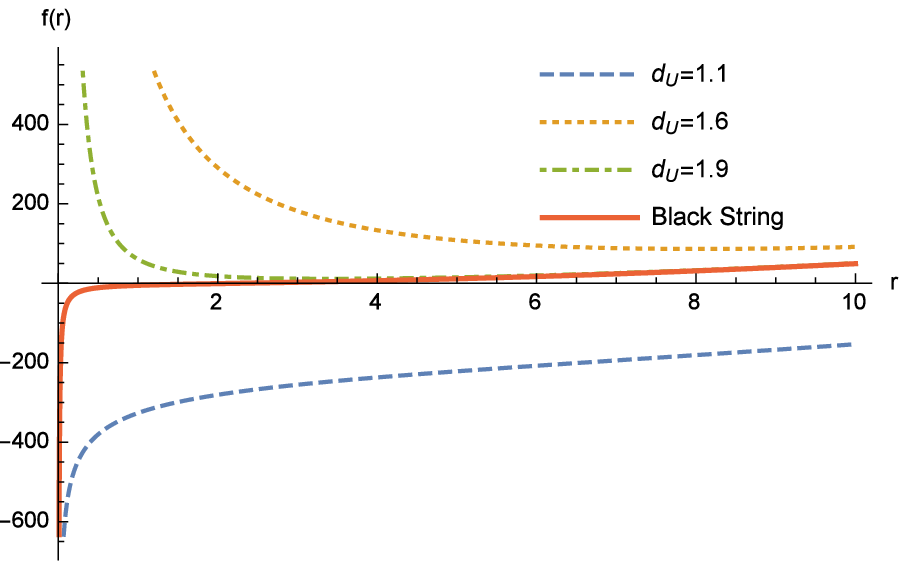}
  \caption{}
 
\end{subfigure}%
\begin{subfigure}{.5\textwidth}
  \centering
  \includegraphics[scale=0.8]{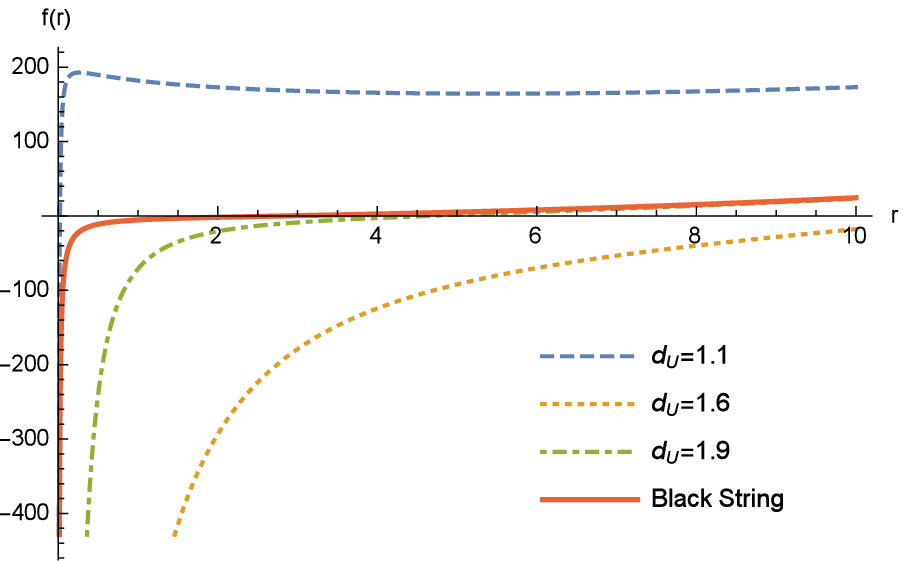}
  \caption{}

\end{subfigure}%
    \caption{Plot of $f(r)$ (scalar case (a) and vector case (b)) for some values of $d_U$ with $\Lambda_{U}=1$, $\alpha=0.5$, $\mu=0.7$, $\kappa=1$ and $\kappa_{*}=10^2$}
    \label{f(r)plot}
\end{figure*}

\begin{eqnarray}\label{29}
    f(r) = \alpha^2r^2 - \frac{4\mu}{\alpha r} \pm \frac{\kappa^2\mu}{2\pi(2d_U-3)}\left(\frac{R}{r}\right)^{2d_U-2},
\end{eqnarray}
with $R$ given by equation (\ref{unradiu}). Note that the unparticle black string solution is divergent for $d_U=3/2$. The plus signal is taken for the scalar unparticle case, while the minus one is for the vector case.

The above solution for the black string in the ungravity scenario recovers the usual black string solution \cite{Lemos:1994xp} in the GD regime, i.e.,
\begin{eqnarray}\label{31}
    f(r)=\left(\alpha^2r^2-\frac{4\mu}{\alpha r}\right).
\end{eqnarray}

Also, it is important to highlight here that the unparticle black string may exhibit naked singularity, violating therefore the cosmic censorship, for the scalar case when $3/2<d_U<2$. 

\section{Thermodynamic properties of the Black String with unparticle corrections}
To make easy the calculations, we can rewrite the equation (\ref{29}) as $f(r) = h(r) + \mu s(r)$, where 
\begin{eqnarray}
    h(r) &=& \alpha^2 r^2, \\
    s(r) &=& \frac{4}{\alpha r}\left[-1 \pm \frac{\alpha r \kappa^2}{8\pi (2d_U-3)}\left(\frac{R}{r}\right)^{2d_U -2}\right].
\end{eqnarray}

Our black string solution in the un-gravity scenario has the horizon curves defined by $f(r_h)=0$, so $\mu = - \frac{h(r_h)}{s(r_h)}$, which gives us
\begin{eqnarray}\label{32}
    \mu= -\frac{\alpha^3r_h^3}{4}\left[-1\pm\frac{\alpha r_h \kappa^2}{8\pi(2d_U-3)}\left(\frac{R}{r_h}\right)^{2d_U-2}\right]^{-1}.
\end{eqnarray}

\begin{figure*}[h!]
    \centering
\begin{subfigure}{.5\textwidth}
  \centering
  \includegraphics[scale=0.8]{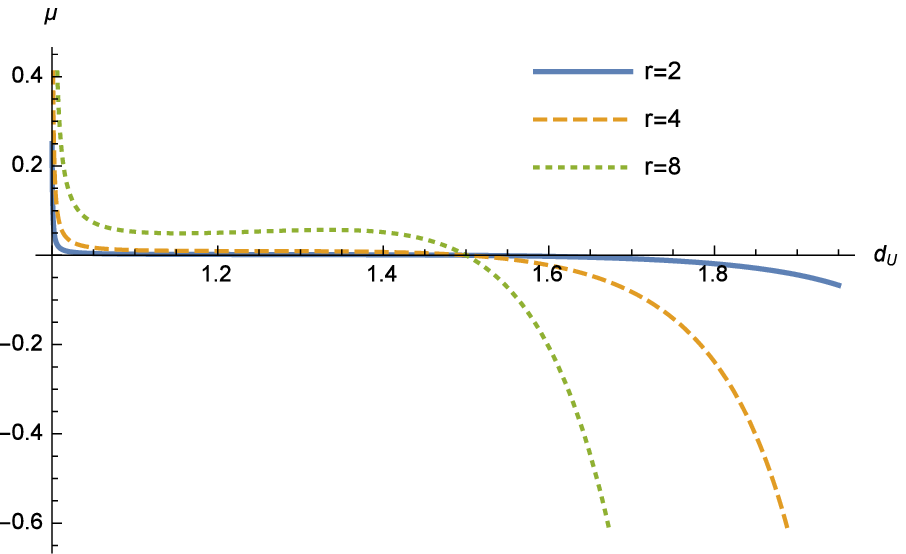}
  \caption{}
 
\end{subfigure}%
\begin{subfigure}{.5\textwidth}
  \centering
  \includegraphics[scale=0.8]{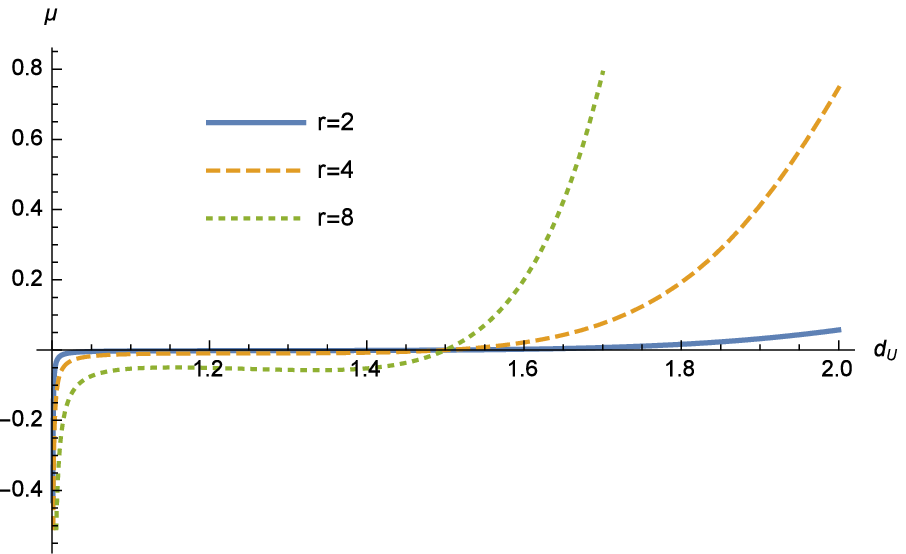}
  \caption{}

\end{subfigure}%
    \caption{Linear mass parameter $\mu$ as a function of $d_U$ for three values of $r$. In (a) we are considering the scalar unparticle case while in (b) we have the vectorial unparticle case. We have considered $\Lambda_{U}=1$, $\alpha=0.5$, $\mu=0.7$, $\kappa=1$ and $\kappa_{*}=10^2$.}
    \label{mu}
\end{figure*}

In the GD regime, the mass parameter $\mu$ reduces to $\mu=\frac{\alpha^3}{4}r_h^3$, which is the usual mass parameter for the black-string in General Relativity. In \cite{Alencar:2018vvb} the authors have found that for an unparticle black hole the mass parameter becomes negative for the vector unparticle case. For the unparticle black string, as we can see in fig.(\ref{mu}), both vector and scalar cases present the possibility of negative mass parameter. For the scalar unparticle case we have a negative mass parameter for values of $3/2<d_U<2$. For the vector unparticle case the mass parameter becomes negative for $1<d_U<3/2$. 

The authors in \cite{Mureika:2010je} does not take into consideration the regions where the mass parameter is negative. However it is known that a negative mass parameter in a black hole solution suggests that it has a nontrivial topology \cite{Mann:1997jb}. Therefore we can expect that these negative mass regions are also related to nontrivial topologies of the the black string.

\begin{figure}[h!]
    \centering
    \includegraphics[scale=0.8]{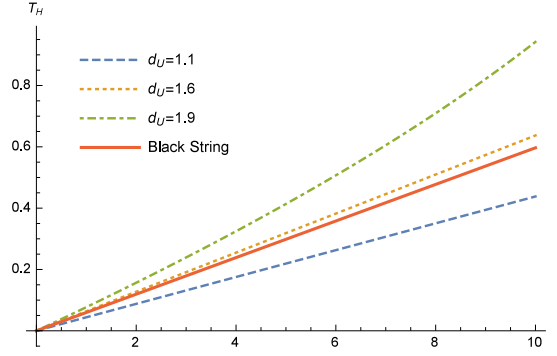}
    \caption{Hawking temperature as a function of $r$ for the scalar unparticle case for three values of $d_U$. We have considered $\Lambda_{U}=1$, $\alpha=0.5$, $\mu=0.7$, $\kappa=1$ and $\kappa_{*}=10^2$.}
    \label{HawkingTemperature}
\end{figure}

In possession of the solution for the static black string in the ungravity scenario given by (\ref{29}), we are able to study the thermodynamics of the black string by computing the Hawking's temperature by means of $T_H=\frac{1}{4\pi}f'(r_h)$, where $r_h$ is the radius of the horizon. Note that in terms of $h(r)$ and $s(r)$ we find 
\begin{eqnarray}\label{temp}
    T_H &=& \frac{1}{4\pi}\left[h'(r_h)+ \mu ~s'(r_h)\right] \nonumber \\
     &=& \frac{s(r_h)}{4\pi}\frac{d}{dr_h}\left(\frac{h(r_h)}{s(r_h)}\right),
\end{eqnarray}
where we used $\mu = -\frac{h(r_h)}{s(r_h)}$. When we substitute the expressions for $h(r_h)$ and $s(r_h)$, we find
\begin{eqnarray}\label{33}
    T_H = \frac{3\alpha^2 r_h}{4\pi}\frac{\left[-1\pm\frac{d_U\alpha r_h\kappa^2}{12\pi(2d_U-3)}\left(\frac{R}{r_h}\right)^{2d_U-2}\right]}{\left[-1\pm\frac{\alpha r_h\kappa^2}{8\pi(2d_U-3)}\left(\frac{R}{r_h}\right)^{2d_U-2}\right]}.
\end{eqnarray}
The usual black string Hawking temperature, i.e. $T_H=\frac{3 \alpha ^2 }{4 \pi }r_h$, is recovered in the GD regime and in UGD regime, we find
\begin{eqnarray}\label{33-1}
    T_H = \frac{\alpha^2 2d_U}{4\pi}r_h,
\end{eqnarray} 
for both GD and UGD regimes. The Hawking temperature for the unparticle black string presents a similar behaviour for both scalar and vector unparticle cases. The behaviour of the Hawking temperature as a function of $r$ for the scalar unparticle case for some values of $d_U$ is presented in figure (\ref{HawkingTemperature}). We can see from figure (\ref{HawkingTemperature}) that when $1<d_U<3/2$ the temperature increases slowly with the radius in comparison with the usual black string temperature. When $3/2<d_U<2$ the unparticle correction promotes an increasing in the temperature in comparison with the black string usual case.  

Note that the temperature is aproximally proportional to $D-1$, where $D$ is the dimension of the spacetime. So, when we confront the equation (\ref{33-1}) with this fact, we can see the fractalization of the event horizon with
\begin{eqnarray}\label{33.2}
 d_H=2d_U-1 ,
\end{eqnarray}
where $d_H = D - 2$. To reinforce the fractalization of the event horizon, we can compute the density of entropy by length $\mathcal{S}$ althrough  $d\mathcal{S} = d\mu/T_H$, which in terms of $h(r)$ and $s(r)$, 
\begin{eqnarray}
    d\mathcal{S} = - \frac{4\pi}{s(r_h)} dr_h,
\end{eqnarray}
where we used equation (\ref{temp}) and again $\mu = - \frac{h(r_h)}{s(r_h)}$. So we find 
\begin{eqnarray}\label{differentialOfEntropy}
    d\mathcal{S} = \frac{\pi\alpha r_h ~dr_h}{\left[-1 \pm \frac{\alpha R \kappa^2}{8\pi (2d_U-3)}\left(\frac{R}{r_h}\right)^{2d_U -3}\right]}.
\end{eqnarray}
Now, we want to find analytically the entropy by length of the black string with unparticle. It can be done by integration of equation (\ref{differentialOfEntropy}), but it is necessary to analyze the regions due to the possible singularity of this equation. Up to now we have discussed the scalar and vector cases simultaneously, but now, for the sake of simplicity we will consider them separately.

\subsection{scalar case:}

Let us begin by considering the case when $d_U<3/2$. In this case we can rewrite the equation (\ref{differentialOfEntropy}) as 
\begin{eqnarray}\label{differentialOfEntropyDSmallerThan3/2}
    d\mathcal{S} = \frac{\pi\alpha r_h ~dr_h}{\left[1 + \frac{\alpha R \kappa^2}{8\pi (3-2d_U )}\left(\frac{R}{r_h}\right)^{2d_U -3}\right]}. 
\end{eqnarray}
So, in case $d_U<3/2$ there is not singularity. By using the identity
\begin{eqnarray}\label{identity}
\int \frac{x dx}{1\pm Ax^\gamma} = \frac{x^2}{2}\hyper{1}{\frac{2}{\gamma}}{1+\frac{2}{\gamma}}{\mp Ax^\gamma} + \text{const.},
\end{eqnarray}
where $\hyper{a}{b}{c}{x}$ is the hypergeometric function and $\gamma>0$, the integration of equation (\ref{differentialOfEntropyDSmallerThan3/2}) can be written as
\begin{eqnarray}\label{entropy1}
    S = \frac{2\pi\alpha {r_h}^2}{4}\hyper{1}{\frac{2}{3-d_U}}{1+\frac{2}{3-d_U}}{- \frac{\alpha R \kappa^2}{8\pi (3-2d_U )}\left(\frac{R}{r_h}\right)^{2d_U -3}}.
\end{eqnarray}
Note that, in GD regime ($R\rightarrow 0$) the argument of hypergeometric funcion is proportional to $R^{2d_U-2}$, which approaches to $0$ when $R \rightarrow 0$. It is, the entropy in GD recovers  $S = \frac{2\pi\alpha {r_H}^2}{4}$ as shall be.

The heat capacity and the free energy of the system are defined by
\begin{eqnarray}
    \label{heat}C_V&=&\frac{d\mu}{dT_H}\\
    \label{free}F&=&\mu-T_H S.
\end{eqnarray}
The heat capacity can be obtained directly from the equations (\ref{32}) and (\ref{33}), which yields
\begin{eqnarray}
    C_V=\frac{8\pi^2(-3+2d_U)\alpha(r_h R)^2\left[12(-3+2d_U)\pi R^2-d_U\kappa^2\alpha R^3\left(\frac{R}{r_h}\right)^{2d_U-3}\right]}{96(3-2d_U)^2\pi^2R^4+8\pi(-9+27d_U-20d_U^2+4d_U^3)\kappa^2r_h^3R^2\alpha\left(\frac{R}{r_h}\right)^{2d_U}+d_U\alpha^2\kappa^4r_h^6\left(\frac{R}{r_h}\right)^{4d_U}}.
\end{eqnarray}
The free energy can also be obtained from (\ref{32}), (\ref{33}) and (\ref{entropy1}), yielding
\begin{eqnarray}
    \nonumber F&=&-\frac{\alpha^3 r_h^3}{4 \alpha  \kappa^2 r_h^{3-2 d_U} R^{2 d_U}+32 \pi  (3-2 d_U) R^2} \left[8 \pi  (2 d_U-3) R^2\right.\\
    &&+\left.\left(\alpha d_U \kappa^2 r_h^{3} \left(\frac{R}{r_h}\right)^{2 d_U}+12 \pi  (3-2 d_U) R^2\right) \, _2F_1\left(1,\frac{2}{3-2 d_U};1+\frac{2}{3-2 d_U};-\frac{\alpha  \kappa^2 r_h^{3-2 d_U} R^{2 d_U-2}}{24 \pi -16 \pi  d_U}\right)\right].
\end{eqnarray}

\begin{figure*}[h!]
    \centering
\begin{subfigure}{.5\textwidth}
  \centering
  \includegraphics[scale=0.8]{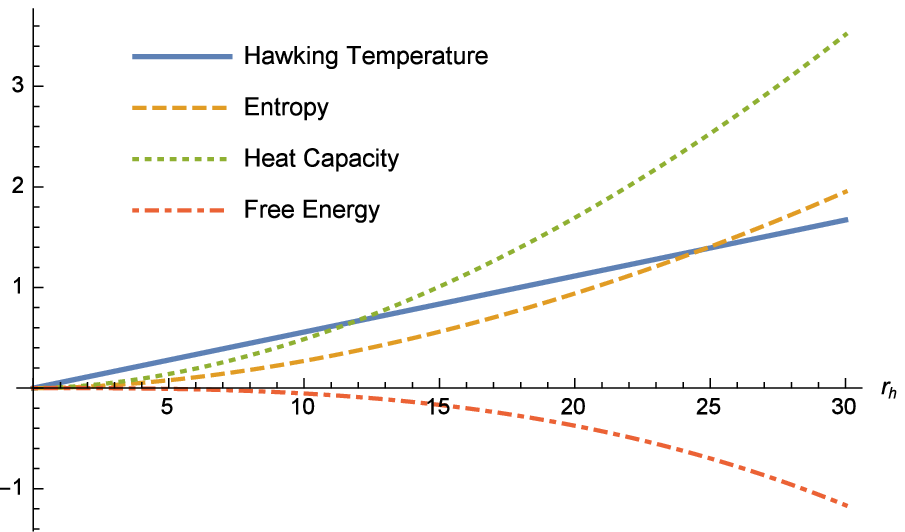}
  \caption{}
 
\end{subfigure}%
\begin{subfigure}{.5\textwidth}
  \centering
  \includegraphics[scale=0.8]{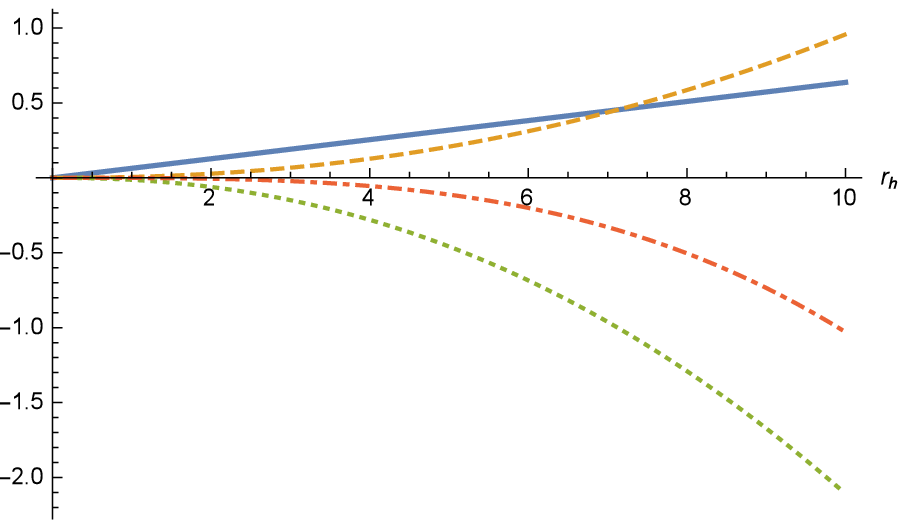}
  \caption{}

\end{subfigure}%
    \caption{Hawking temperature, entropy, heat capacity and free energy for $d_U=1.4$ (a) and for $d_U=1.6$ (b)}
    \label{therm_quant_1}
\end{figure*}

The behaviour of the Hawking temperature, entropy, heat capacity and free energy for $d_U<3/2$ is depicted in figure (\ref{therm_quant_1}).

Now, we need to investigate the case $d_U>3/2$. In this case, there is a singularity in equation (\ref{differentialOfEntropy}) when 
\begin{eqnarray}\label{singularityForr_H}
    r_0 = \left[\frac{\alpha R \kappa^2}{8\pi (2d_U-3)}\right]^{\frac{1}{2d_U -3}}R,
\end{eqnarray}
It is easy to see that we have two disjunte regions and when $r_h >r_0$ $\left(r_h <r_0\right)$  the differential of entropy by length, equation (\ref{differentialOfEntropy}), approaches infinity (minus infinity). First, let us compute the entropy for $r_h >r_0$.We can rewrite equation (\ref{differentialOfEntropy}) as
\begin{eqnarray}\label{differentialOfentropyOfDLargerThan3/2}
    d\mathcal{S} = \frac{\alpha \pi r_hdr_h}{\left[1 -\left(\frac{r_0}{r_h}\right)^{2d_U-3}\right]}.
\end{eqnarray}
We can make a changing of variable $z = (r_o/r_h)^{2d_U-3}$, equation (\ref{differentialOfentropyOfDLargerThan3/2}) becomes
\begin{eqnarray}\label{differentialOfentropyOfDLargerThan3/2-withz}
    d\mathcal{S} = -\frac{\alpha \pi{r_0}^2}{(2d_U-3)}~\frac{ z^{-\left(\frac{2d_U-1}{2d_U-3}\right)}dz}{1 -z}.
\end{eqnarray}
We can use the identity 
\begin{eqnarray}\label{identity2}
    \int \frac{x^{\pm c}dx}{1-x} =  \frac{x^{1\pm c}}{1\pm c}~\hyper{1}{1\pm c}{1+(1\pm c)}{x} + \text{const.},
\end{eqnarray}
where $c>0$. So, when we integrate the equation (\ref{differentialOfentropyOfDLargerThan3/2}), we find
\begin{eqnarray}\label{entropyOfDLargerThan3/2-1}
\mathcal{S} = \frac{2\pi \alpha r_h^2}{4}~\hyper{1}{\frac{2}{3-2d_U}}{1+\frac{2}{3-2d_U}}{\left(\frac{r_0}{r_h}\right)^{2d_U-3}}.
\end{eqnarray}
Note that equation (\ref{entropyOfDLargerThan3/2-1}) recovers $\mathcal{S} = \frac{2\pi \alpha r_h^2}{4}$ in GD regime. The ultimate case is when $r_h <r_0$. We can rewrite equation (\ref{differentialOfEntropy}) as
\begin{eqnarray}\label{changingOfVariable2} 
    d\mathcal{S} =-\alpha \pi \left(\frac{r_h}{r_0}\right)^{2d_U-3}\frac{ r_h dr_h}{\left[1-\left(\frac{r_h}{r_0}\right)^{2d_U-3}\right]}.
\end{eqnarray}
When we make a changing of variable $z = (r_h/r_0)^{2d_U-3}$, equation (\ref{changingOfVariable2}) becomes
\begin{eqnarray}\label{changingOfVariable2-3} 
    d\mathcal{S} =-\frac{\alpha \pi {r_0}^2}{(2d_U-3)}~ \frac{ z^{\frac{2}{2d_U-3}} dz}{1-z}.
\end{eqnarray}

So, when we integrate the equation (\ref{changingOfVariable2-3}) by using equation (\ref{identity2}), we find
\begin{eqnarray}\label{entropyOfDLargerThan3/2-2}
    \mathcal{S} = - \frac{\pi \alpha {r_0}^2}{(2d_U-1)}\left(\frac{r_h}{r_0}\right)^{2d_U-1} \hyper{1}{\frac{2d_U-1}{2d_U-3}}{\frac{4(d_U-1)}{2d_U-3}}{\left(\frac{r_h}{r_0}\right)^{2d_U-3}}
\end{eqnarray}
Note that in UGD regime the depencence of $r_h$ in equation (\ref{entropyOfDLargerThan3/2-2}) becomes $r_h^{d_h} = r_h^{2d_U-1}$. Then $d_h = 2d_U -1 $, which reinforces the fractalization and is in agreement with equation (\ref{33-1}). 

The heat capacity can be obtained straightforwardly from the equations (\ref{32}) and (\ref{33}), yielding
\begin{eqnarray}
    C_v=\frac{8 \pi ^2 \alpha  (2 d_U-3) R^2 r_h^{2 d_U+2} \left(12 \pi  (2 d_U-3) R^2 r_h^{2 d_U}-\alpha  d \kappa^2 r_h^3 R^{2 d_U}\right)}{\alpha ^2 d \kappa^4 r_h^6 R^{4 d_U}+8 \pi  \alpha  (d_U-3) (2 d_U-3) (2 d_U-1) \kappa^2 r_h^3 R^2 (r_h R)^{2 d_U}+96 \pi ^2 (3-2 d_U)^2 R^4 r_h^{4 d_U}}.
\end{eqnarray}

The free energy can also be obtained from (\ref{32}), (\ref{33}) and (\ref{entropy1}), from which follows that

\begin{eqnarray}
    F=\frac{\alpha ^3 r_h \left(r_0^2 B_{\left(\frac{r}{r_0}\right)^{2 d_U-3}}\left(1+\frac{2}{2 d_U-3},0\right) \left(\alpha  d \kappa^2 r_h^3 R^{2 d_U}-12 \pi  (2 d_U-3) R^2 r_h^{2 d_U}\right)+4 \pi  (3-2 d_U)^2 R^2 r_h^{2 d_U+2}\right)}{2 (2 d_U-3) \left(8 \pi  (2 d_U-3) R^2 r_h^{2 d}-\alpha  \kappa^2 r_h^3 R^{2 d_U}\right)},
\end{eqnarray}
where $B_{z}\left(a,b\right)$ is the incomplete Euler Beta function.

As it is widely known, the thermodynamical stability of black holes (black strings for our case) is directly related to the sign of the heat capacity. A positive heat capacity indicates that the system is thermodynamically stable, while its negativity imply a thermodynamical instability. As we can see in figure (\ref{therm_quant_1}), the value of the $d_U$ controls the black string stability, being the black string stable in the ungravity regime for $d_U<3/2$ and unstable for $d_U>3/2$.

\subsection{vector case:}

Let us consider the case when $d_U>3/2$, which for the vector case avoid the singularity, so that the equation (\ref{differentialOfEntropy}) can be written as
\begin{eqnarray}\label{dsvector}
    d\mathcal{S}=\frac{\pi\alpha r_hdr_h}{\left[1+\frac{\alpha\kappa^2 R}{8\pi(2d_U-3)}\left(\frac{R}{r_h}\right)^{2d_U-3}\right]}.
\end{eqnarray}
We can clearly see that, in virtue of the similarity between the expressions (\ref{dsvector}) and (\ref{differentialOfEntropyDSmallerThan3/2}), the same analysis performed for the scalar case when $d_U<3/2$ can be done for the vector case when $d_U>3/2$. Also the results for the scalar case when $d_U>3/2$ are identical to those of the vector case when $d_U<3/2$. 

Therefore, the thermodynamic quantities depicted in figure (\ref{therm_quant_1}a) for the scalar case when $d_U<3/2$ also describe the thermodynamic quantities for the vector case when $d_U>3/2$. Likewise, the thermodynamic quantities depicted in figure (\ref{therm_quant_1}b) for the scalar case when $d_U>3/2$ are also describing the thermodynamic quantities for the vector case when $d_U<3/2$.

\section{Conclusion}\label{Sec-6}

In this paper we study the unparticle corrections for cosmic and black strings. Initially we have discussed some general features about the unparticle physics from which we obtained a general expression for the energy density for a string. We could identify a proper length scale $R$ associated with the unparticle scenario. such length scale define two regimes, namely, a gravity dominated regime and an ungravity dominated regime. We find that the unparticle corrects the a string source by
\begin{eqnarray}\label{source}
    \rho(r) = \frac{\mu}{2\pi}\left[\frac{\delta(r)}{r}+\frac{R^{2d_U-2}}{r^{2d_U}}\right],
\end{eqnarray}

With the above source, we first have considered an static cosmic string ansatz from which we obtain the solution in terms of first and second kind Bessel functions. The structure of the cosmic string spacetime in the unparticle scenario was studied by means of the Ricci scalar.   

 However, in order to study  fractalization, we must consider the black hole analogous with cylindrical symmetry. In order to get this, first we obtained the solution for black string in the unparticle scenario directly from the Einstein field equations. The solutions were studied for both scalar and vector unparticle case, and in both cases the solution is divergent when $d_U=3/2$. In both cases the unparticle black string solution recovers the usual black string solution in the gravity dominated regime, as expected. Also, it is important to highlight that, in the black string, all the cases considered  may exhibit a naked singularity depending on the range of $d_U$. For the unparticle black hole this was a feature present only for the vector case\cite{Alencar:2018vvb}.

The thermodynamic properties of the black string with the unparticle corrections were also addressed. We could verify that either in scalar unparticle case or in vector case, there are regions with negative mass parameter. For the scalar case, the mass becomes negative when $3/2<d_U<2$, while for the vector case the mass becomes negative when $1<d_U<3/2$. As pointed out in \cite{Mann:1997jb}, black holes (black strings, for our case) with negative mass can exist and present non-trivial topology. By investigating the Hawking temperature we could see that both scalar and vector cases present a very similar behaviour. When $1<d_U<3/2$ the temperature increases slowly with the radius in comparison with the usual black string temperature. When $3/2<d_U<2$ the unparticle correction promotes an increasing in the temperature in comparison with the black string usual case. 

With the above solution we analyze if, similarly to the unparticle black hole cases, we can have a fractalization. We could identify that for the black string in the UGD regime we have an effective dimension given by $d_H=2d_U-1$. We should point that, for the static and rotating black hole, the fractal dimension is given by $d_H=2d_U$. The fractalization can also be obtained from the entropy, which was analytically obtained for the scalar and vector cases. For both cases it is given by the Hypergeometric confluent functions. In the UGD regime we find again a fractal dimension given by  $d_H=2d_U-1$.  Entropy and other thermodynamical quantities also give important information about the stability of the system. We perform our analysis separately for $d_U>3/2$ and $d_U<3/2$ in order to avoid singularities. We compute the specific heat and free energy for the scalar case and plots are depicted in figure (\ref{therm_quant_1}) for a particular choice of parameters. We could see that the value of the $d_U$ controls the black string stability in the ungravity scenario, being the black string stable for $d_U<3/2$ and unstable for $d_U>3/2$. 

Finally we point that the source \ref{source} will give modified light deflection and could be found in the present era of high precision astrophysical measurements. It would also be interesting to study if the rotating case give the same fractal dimension, as in the black hole case. In fact, the study of fractal dimension in unparticle can be done to many other objects. A more ambitious project would be to show that horizon fractalization is always present in unparticle gravity.

\vfill
\section*{Acknowledgement}
The authors would like to thanks Alexandra Elbakyan and sci-hub, for removing all barriers
in the way of science. We acknowledge the financial support provided by the Coordenação de Aperfeiçoamento de Pessoal de Nível Superior or (CAPES), the Conselho Nacional de Desenvolvimento Científico e Tecnológico (CNPq) and Fundaçao Cearense de Apoio ao Desenvolvimento Científico e
Tecnológico (FUNCAP) through PRONEM PNE0112- 00085.01.00/16.

\end{document}